\documentclass[aps,prd,onecolumn,12pt,tightenlines,notitlepage,groupedaddress,longbibliography]{revtex4-2}

\usepackage{amsmath,amssymb,amsfonts}
\usepackage[T1]{fontenc}
\usepackage{graphicx}
\usepackage{array}
\usepackage{hyperref}

\begin{document}


\title{No single unification theory of everything}


\author{Wanpeng Tan}
\email[]{wtan@nd.edu}
\affiliation{Department of Physics, Institute for Structure and Nuclear Astrophysics (ISNAP), and Joint Institute for Nuclear Astrophysics - Center for the Evolution of Elements (JINA-CEE), University of Notre Dame, Notre Dame, Indiana 46556, USA}


\date{\today}

\begin{abstract}
In light of G\"{o}del's undecidability results (incomplete theorems) for math, quantum indeterminism indicates that physics and the Universe may be indeterministic, incomplete, and open in nature, and therefore demand no single unification theory of everything. The Universe is dynamic and so are the underlying physical models and spacetime. As the 4-d spacetime evolves dimension by dimension in the early universe, consistent yet different models emerge one by one with different sets of particles and interactions. A new set of first principles are proposed for building such models with new understanding of supersymmetry, mirror symmetry, and the dynamic phase transition mechanism – spontaneous symmetry breaking. Under this framework, we demonstrate that different models with no theory of everything operate in a hierarchical yet consistent way at different phases or scenarios of the Universe. In particular, the arrow of time is naturally explained and the Standard Model of physics is elegantly extended to time zero of the Universe.
\end{abstract}


\maketitle

\section{Introduction\label{intro}}

G\"{o}del's incomplete theorems, proved in 1931, demonstrate that any consistent math system with elementary arithmetic is incomplete, i.e., there exist statements that can not be proved or disproved in this system. Such undecidability results have profoundly impacted on mathematics and philosophy. In particular, it may have delivered a crushing blow to Hilbert's program for unifying all mathematics under a finite complete set of axioms.

However, scientists, especially physicists do not easily give up dreams like unification of their field. Einstein, as the best scientist of modern times, had openly presented his failed effort of pursuing a unification theory of physics based on his relativity theory until his death in 1955. Other physicists, taking a different route via quantum theory, had essentially unified all three gauge interactions (i.e., electromagnetic, nuclear weak and strong forces) and eventually established the so-called Standard Model of particle physics in 1970s. It took about half an century from many physicists' efforts since the establishment of quantum mechanics in 1920s.

The Standard Model, as the best known quantum theory, is probably the most influential achievement in modern physics. It has led to so many critical applications in our daily life and has been tested at amazingly high precision. Yet most physicists are not satisfied with this theory. The gauge symmetry group $SU_c(3)\times SU_L(2)\times U_Y(1)$ of the three united fundamental forces looks like a patchwork. There are many ad hoc parameters and puzzles within the theory itself like three generations of elementary particles and nature of neutrinos. More curiously, it does not seem to be compatible with the best known gravity theory - Einstein's general relativity that is the other pillar of modern physics.

Over another half century, many physicists have attempted some sort of grand unification theory, also known as quantum gravity or theory of everything, trying to unify all fundamental forces. One of the most pursued unification schemes is called string theory. In particular, an idea for a new symmetry called supersymmetry (SUSY) that relates fermions (particles with half-integer spins) to bosons (particles with integer spins) has been proposed. However, the idea seems to be in a crisis as the Large Hadron Collider has found no evidence of supersymmetric particles so far. None of these attempts has produced any credible and consistent understanding of various outstanding puzzles in physics (e.g., dark energy, dark matter, and matter-antimatter asymmetry of the Universe). Both our theoretical and experimental efforts in search of new physics beyond the Standard Model, other than probably the effort of neutrino oscillations, have failed miserably, at least so far. For example, despite tremendous technology advancements over the past decades, dark matter search has not detected any imagined particles.

This should make us ponder if we are going in the correct direction. Are we paying the price for ignoring the advice from G\"{o}del's incomplete theorems? In the following, a new way presenting a dynamic hierarchy of physics and the Universe without assuming a single unification theory is discussed based on the most recent works related to mirror matter theory \cite{tan2019,tan2019a,tan2019b,tan2019c,tan2019d,tan2019e,tan2020}. We will start in the next section to show how the new principles of dynamic symmetries and mechanisms for new physics would be proposed under evolving spacetime dimensions.

\section{New principles of physics\label{prin}}

In review of the two pillars of modern physics - quantum theory and general relativity, we could see one of the biggest discrepancies between them: indeterminism in quantum theory and determinism in general relativity. The reason behind the controversies of quantum foundations such as measurement and nonlocality issues is that quantum theory is intrinsically indeterministic, i.e., probability or more exactly amplitude of probability is built in the heart of quantum theory. On the contrary, general relativity and other classical theories are fundamentally deterministic and tend to give a ``complete'' picture of the world that is fully determined upon the knowledge of initial conditions. In accord with G\"{o}del's incomplete theorems, quantum theory does not provide a ``complete'' or decidable picture for various cases like Schr\"{o}dinger's cat.

This indicates that indeterminism of our world comes from the quantum nature or our quantum world is indeterministic in nature by analog to G\"{o}del's undecidability in math. But it does not mean that our world has to be chaotic or follows no orders at all. On the contrary, certain orders are emergent as demonstrated by a hierarchy of models that will be discussed below for different phases of the Universe. We can still provide a self-consistent description within a particular domain of physics or for a certain stage of the Universe and maybe even establish consistent connections between theories in different domains. However, we will not be able to build an all-inclusive type unification theory in physics. In other words, we can not deduce all the details of theories of various domains from a single universal theory, not even in principle. New laws or phenomena will keep popping up when we broaden our horizon or cross from one domain to another.

If quantum indeterminism is fundamental in nature, then determinism must be an emergent phenomenon, most likely, from the average effect of quantum fluctuations. In particular, general relativity is likely just a mean field theory emergent from the inflation of quantum spacetime \cite{tan2020}. Gravity can be considered as the smooth mean geometry after averaging out quantum fluctuations of extended or inflated spacetime and it serves as the mean background spacetime for quantum particles. Concisely, determinism of general relativity is emergent from the underlying indeterministic quantum theory.

An immediate implication of quantum indeterminism is that our world has to be dynamic with phase transitions. It is hard to imagine that a static universe is not deterministic unless one assumes different laws at different locations, which, however, defeats the static assumption for spacetime. The best known dynamic mechanism of phase transitions in quantum theory is called spontaneous symmetry breaking \cite{nambu1961}. The Higgs mechanism in particle physics that gives mass to particles is based on this approach. However, Higgs and other similar scalars are likely bound states of fermions as demonstrated in the application of this mechanism for staged quark condensation or electroweak / QCD phase transitions \cite{tan2019c,tan2019e}. That is exactly what we need to transform one theory to another in a dynamic process.

Now we need to find out how symmetry works given a spacetime configuration. Our Universe's spacetime can be described mathematically by a Riemannian manifold. The holonomy group of an n-dimensional Riemannian manifold as a maximal symmetry of distance-preserving is the orthogonal group $O(n)$ that defines the local / gauge symmetry for the interactions. Note that $O(n) \sim SO(n)\times Z_2$ where $SO(n)$ is the special orthogonal or rotation group and $Z_2$ is a parity-like discrete symmetry containing two group elements of $\{1,-1\}$. Here we will call this $Z_2$ group generalized mirror symmetry and its concrete meaning under different spacetime dimensions will be discussed later.

The idea of mirror symmetry was originated from the seminal work on parity violation by Lee and Yang \cite{lee1956}. Could the elementary particles have an almost identical mirror copy instead of an often-assumed supersymmetric copy? It is conceivable that there exist two sectors of particles sharing the same gravity but governed by two separate gauge groups under 4-d spacetime. Some early works on mirror matter theory had discussed interesting perspectives in terms of mainly cosmology \cite{blinnikov1983,kolb1985,hodges1993}. Other attempts to introduce feeble interactions between the two sectors might be going in the wrong way \cite{berezhiani2006,cui2012,foot2014}. Latest works \cite{tan2019,tan2019a,tan2019b,tan2019c,tan2019d,tan2019e,tan2020} by keeping only the essence of mirror matter theory may indeed lead us to new physics beyond the Standard Model we have all been looking for.

Considering the mirror symmetry's new role, our understanding of supersymmetry (SUSY) has to be changed. It is the mirror symmetry that demands another copy of particles instead of supersymmetry. SUSY might have already been built in a certain way in the particle zoo and therefore could be used to set further constraints in physical laws. Nambu demonstrated a different SUSY principle as he called it quasi-SUSY \cite{nambu1988,*nambu1988a}. He observed the matching of degrees of freedom (DoF) between fermions and bosons in many models including a special case considering only one generation of fermions in the Standard Model. Such an idea definitely sheds new light on SUSY in building the extended Standard Model with mirror matter \cite{tan2019e,tan2020}.

Under the new SUSY principle, SUSY-partners of the Standard Model particles have already been built in via DoF matching between matter fermions and gauge bosons. SUSY is exactly restored above the energy scale ($\sim 10^2$ GeV) of the electroweak phase transition while becoming pseudo-SUSY between bosons and fermions that are no longer massless after spontaneous symmetry breaking and staged quark condensation \cite{tan2019e,tan2020}. Therefore, SUSY with new understanding evidently becomes very powerful in constraining the gauge group and particle generations in order to keep the DoF balance.

In addition, SUSY requires different sets of particle fields under different spacetime dimensions. For extended superspace ($x^{\mu}$,$\theta^{\alpha}$,$\bar{\theta}^{\dot{\beta}}$) where $x^{\mu}$ are spacetime coordinates and $\theta^{\alpha}$ and $\bar{\theta}^{\dot{\beta}}$ are corresponding anticommuting Grassmannian coordinates, a general scalar function $F(x^{\mu},\theta^{\alpha},\bar{\theta}^{\dot{\beta}})$ can be expanded in finite number of terms up to $\theta^2\bar{\theta}^2$ for defining the fields allowed in 4-d spacetime \cite{weinberg1995}. In 1-d time, $\theta$ and $\bar{\theta}$ do not exist so that the general function $F = \varphi(t)$ is trivial, i.e., only a scalar field $\varphi$ (spin=0) can exist. In 2-d spacetime, $F=\phi(x,t)+\theta\lambda(x,t)+\bar{\theta}\bar{\lambda}(x,t)+\theta\sigma_{\mu}\bar{\theta}A^{\mu}(x,t)$ where more fields like Majorana fermion $\lambda$ and gauge boson $A^{\mu}$ are allowed in addition to a new complex scalar $\phi$. In particular, under varied spacetime dimensions, the scalar fields provide the spontaneous symmetry breaking mechanism and new mass scales for the dynamic evolution of the Universe.

Under such a dynamic framework, SUSY essentially defines different types of gauge interactions and particle species in different spacetime dimensions. In other words, physics becomes completely different under different spacetime configurations.

The Feynman path integration formalism provides a natural way to construct the proper Lagrangian and action for a given system. Under this framework, the probability amplitude of a physical system can be obtained as a coherent sum of all possible configurations weighted by a phase factor of the action of each configuration,
\begin{equation}
A = \sum_{\text{configurations}} \exp(i\mathcal{S}/\hbar)
\end{equation}
where $|A|^2$ determines the probability and the action $\mathcal{S}$ is determined by an integration of Lagrangian over D-dimensional flat spacetime,
\begin{equation}
\mathcal{S} = \int d^{D}x \mathcal{L}.
\end{equation}
Each configuration is associated with its own action or Lagrangian and all configurations can contribute to the probability amplitude. But most of them tend to vanish or cancel out each other. Here we postulate the principles for the action and Lagrangian of the most probable configuration:
\begin{enumerate}
\item It has to be maximally symmetric.
\item It has to be finite or renormalizable.
\item It has to be complete to include all possible terms.
\item Spontaneous symmetry breaking (SSB) provides the dynamic mechanism.
\end{enumerate}

These principles ensure that the Lagrangian should include all possible renormalizable maximal-symmetry-obeying dimension-D (or dimension-4 for 4-d spacetime) terms. At least some of these are commonly applied in constructing Lagrangians for many physical systems. Finiteness essentially selects renormalizable dimension-4 terms as non-renormalizable terms will cause infinity in the action making its contribution to the amplitude vanish. In a given system like our universe, the maximal symmetry postulate requires that all the terms in Lagrangian have to obey Lorentz (for base manifold - extended spacetime), gauge (for compacified and fiber spaces), and other (e.g., SUSY and mirror) symmetries. It has to be maximally symmetric as any asymmetric constructions of Lagrangian tend to cancel out with corresponding asymmetric counterparts. Last but not least, SSB is essential to provide the dynamics for the emergence of hierarchical supersymmetric mirror models.

\section{Hierarchical dynamics of physics and the Universe\label{dyn}}

\begin{figure}
\includegraphics[scale=0.75]{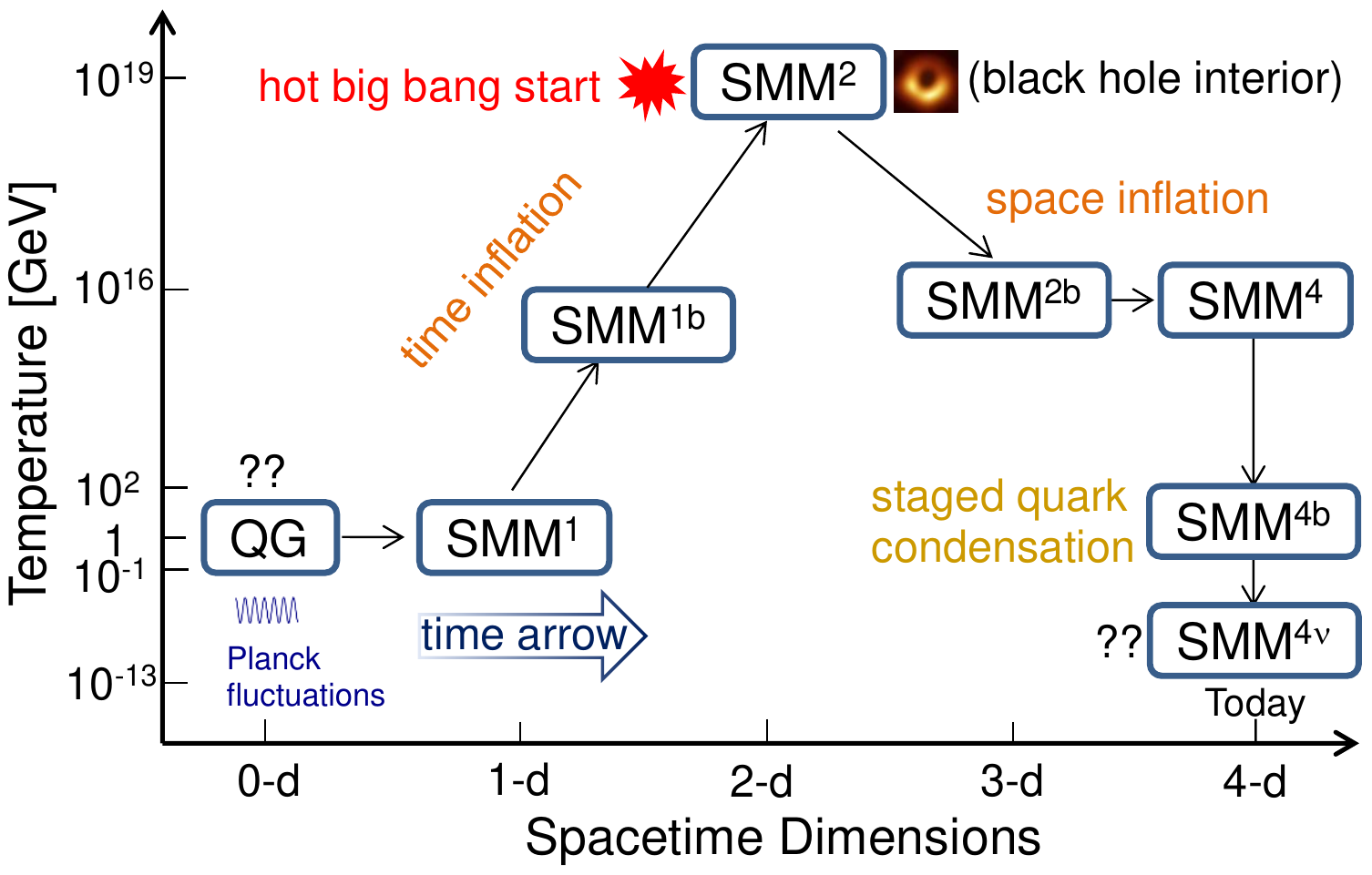}
\caption{\label{fig:smm} The schematic diagram (not to scale) is shown for the dynamic hierarchy of quantum gravity (QG) and supersymmetric mirror models (SMM) at various phases of the Universe and spacetime. The superscript number in model name SMM denotes the number of spacetime dimensions while superscript `b' indicates that the model is for the corresponding spontaneous symmetry breaking process.}
\end{figure}

We assume the zero-dimensional spacetime is of its baby or planckian size in nature which could be described by quantum gravity (QG) proposals like loop quantum gravity \cite{rovelli2014}. By zero-dimension, we mean that all four spacetime dimensions are unextended and there are no particle fields at all but spacetime itself. When spacetime dimensions start to get extended or inflated one by one (first time and then space dimensions), supersymmetric mirror models (SMM) will come into play accordingly with different set of particles and gauge interactions each time in a hierarchical way as shown in Fig. \ref{fig:smm}. We will discuss these models briefly here and see Refs. \cite{tan2019e,tan2020} and \hyperref[app]{Appendix} for more technical details. Some key properties of the models are summarized in Table \ref{tab:1}.

First we look at the case of 1-d time by applying the first principles. In 1-d time, only a true real scalar field $\varphi(t)$ can exist. The Lagrangian for SMM$^1$ can be written simply as,
\begin{equation}\label{eq:l1d}
\mathcal{L} = \frac{1}{2}\dot{\varphi}^2
\end{equation}
for a massless $\varphi$ field.

\begin{figure}
\includegraphics[scale=0.45]{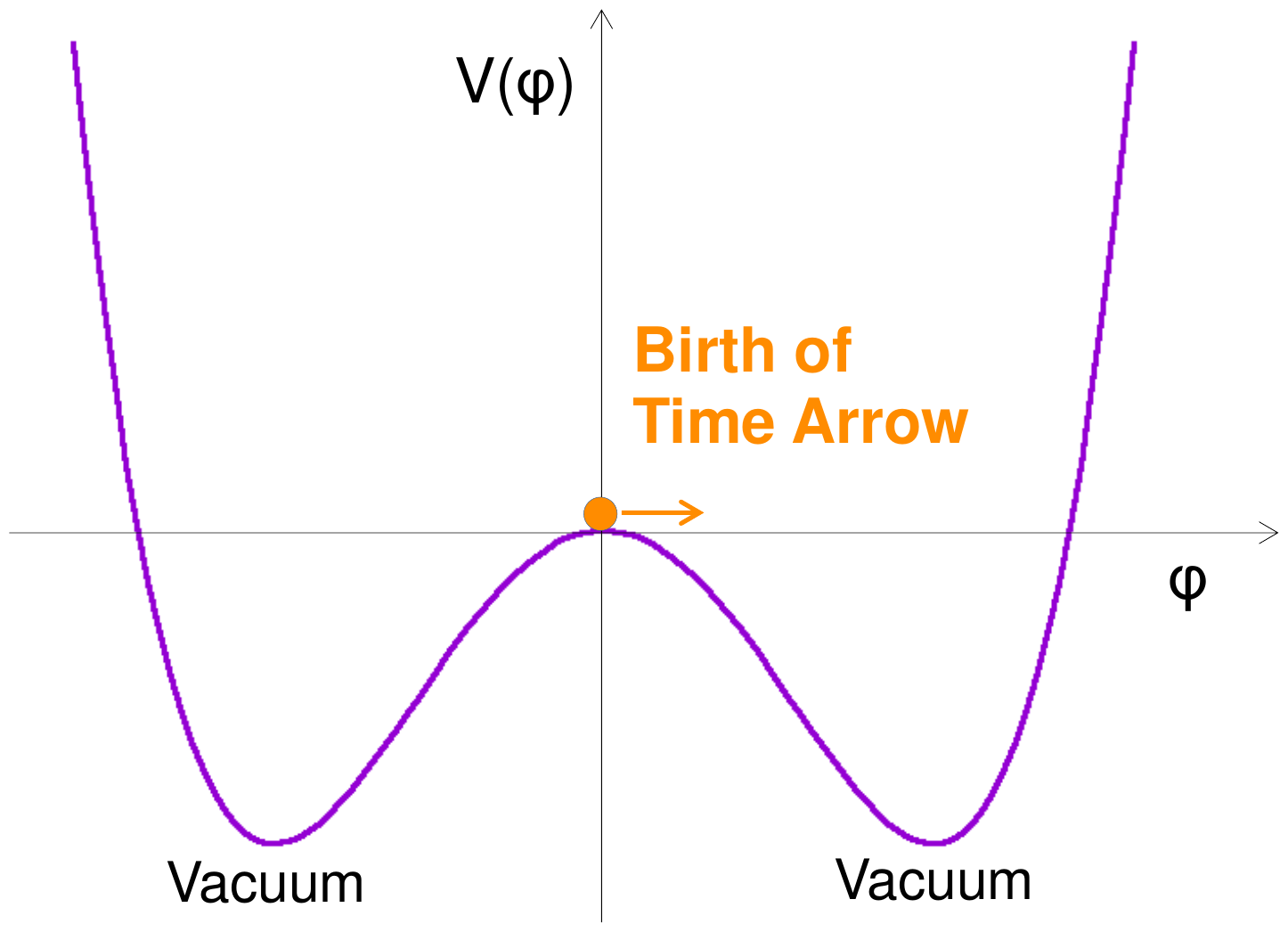}
\caption{\label{fig:hat} The potential of the scalar ``timeron'' field $\varphi$ is shown to demonstrate the time inflation and the birth of the time arrow. Adapted from Ref. \cite{tan2020}}.
\end{figure}

The holonomy group of a 1-d Riemannian manifold is $O(1) = Z_2$, which is actually the time reversal symmetry at this stage of the Universe. To break this time reversal $Z_2$ symmetry, we can apply the spontaneous symmetry breaking mechanism by introducing a potential term $V(\varphi) = -m^2\varphi^2/2+\varphi^4/8$ for the scalar $\phi$ to acquire mass in SMM$^{1b}$. The time reversal symmetry is then broken as the scalar $\varphi$ rolls down its potential towards one of the two emerging true vacuum configurations leading to the emergence of arrow of the time as shown in Fig. (\ref{fig:hat}). This provides the time-like inflation to make time smooth and causal. Therefore, the scalar field $\varphi$ serves as a time inflaton or ``timeron'' to start the timer of the universe. Original fluctuations at the Planck scale are so chaotic that the original baby-sized spacetime is by no means smooth or causal. But this time inflation process can ensure a causally smooth time dimension afterwards.

One likely way to ensure a slow roll of the scalar field $\varphi$ is by considering that the strength of the gravitational force may be restored to its maximum in the beginning, in other words, the original gravitational constant $G_0$ may be $10^{38}$ times larger than its current value $G$. It is conceivable that the gravity may get weakened as spacetime is exponentially extended. After the inflation, the emerging mass of $\varphi$ should be $\sim 1/\sqrt{G} \sim 10^{19}$ GeV, i.e., the Planck mass corresponding to the weakened gravity due to inflated spacetime. At the end of the time inflation (i.e., settled in the new vacuum), ``timeron'' $\varphi$ will decay into new particles of another phase and start the hot big bang at $\sim 10^{19}$ GeV as shown in Fig. (\ref{fig:smm}).

\begin{table}
\caption{Properties of supersymmetric mirror models (SMM) associated with different spacetime dimensions from 1-d to 4-d are listed. SSB denotes spontaneous symmetry breaking. SM represents particles of the Standard Model in the ordinary sector while SM$'$ stand for the copy of particles in the mirror sector. \label{tab:1}}
\begin{ruledtabular}
\begin{tabular}{l l c c c c c}
model & spacetime & T [GeV] & particle fields & SUSY DoF & $Z_2$ symmetry & gauge group\\
\hline
SMM$^1$ & 1-d time & ? & $\varphi(t)$ & N/A & time-reversal & N/A \\
SMM$^{1b}$ & 1-d $\rightarrow$ 2-d & ?-$10^{19}$ & $\varphi(t)$ & \multicolumn{3}{l}{$Z_2$ breaking $\rightarrow$ arrow of time and hot big bang} \\
SMM$^{2}$ & 2-d & $10^{19}\!-\!10^{16}$ & $\lambda(x,\!t)$,$A_{\mu}(x,\!t)$ & $n_b=n_f=2$ & chiral $Z_2$ & $U(1)$\\
SMM$^{2b}$ & 2-d $\rightarrow$ 4-d & $\sim 10^{16}$ & $\phi(x,t)$,$\lambda(x,t)$ & \multicolumn{3}{l}{$Z_2$ breaking $\rightarrow$ double space inflation} \\
SMM$^{4}$ & 4-d & $10^{16}\!-10^{2}$ & $\psi$,$A_{\mu}$,$\psi'$,$A'_{\mu}$  & $n_b=n_f=96$ & mirror $Z_2$ & 
\begin{tabular}{@{}c@{}} $U_f(6)\times SU_c(3)$\\ $\times SU_w(2)\times U_Y(1)$\\ \& mirror copy \end{tabular} \\
SMM$^{4b}$ & 4-d & $10^{2}-0.1$ & + SSB scalars & \multicolumn{3}{l}{$Z_2$ breaking $\rightarrow$ staged quark condensation} \\
SMM$^{4b}$ & 4-d now & $<0.1$ & SM and SM$'$ & 
\begin{tabular}{@{}c@{}} pseudo-SUSY \\ $n_b=n_f=90$ \end{tabular}
& broken $Z_2$ & 
\begin{tabular}{@{}c@{}} $SU_c(3)\!\times\! SU_L(2)\!\times\! U_Y(1)$ \\ \& mirror copy \end{tabular}
\end{tabular} 
\end{ruledtabular}
\end{table}

Now we can consider further evolution of spacetime, that is, 1-d space also starts to grow from its baby size. In 2-d spacetime, the holonomy group is $O(1,1) \sim U(1) \times Z_2$ where the first gauge group $U(1)$ (similar to electromagnetic gauge interaction), supersymmetry, and Majorana fermion $\lambda$ (which is its own anti-particle) are born at this stage. Both the Majorana fermion and the gauge boson are massless with two components or degrees of freedom each (i.e., $n_b=n_f=2$ as shown in Table \ref{tab:1}). They form the simplest abelian gauge SUSY multiplet $(1,1/2)$ with the on-shell Lagrangian (SMM$^2$) shown in \hyperref[app:2]{Appendix}. At the end of the time inflation, the scalar ``timeron'' $\varphi$ with energy of about $10^{19}$ GeV will decay into these two particles when the 2-d spacetime is extended. The $Z_2$ symmetry is just the chiral symmetry between left- and right-handed Majorana fermions at this stage.

Again, spontaneous symmetry breaking will occur when Majorana fermions start to condense into two scalars of $\phi$ and $\phi'$. The $U(1)$ gauge symmetry is broken and the new Lagrangian (SMM$^{2b}$) at the new emerging energy scale of $10^{16}$ GeV shows the symmetry of $N=1$ pseudo-SUSY multiplet $(\frac{1}{2},0)$. Both $\phi$ and $\phi'$ present a Higgs-like dynamic mechanism by acquiring similar yet different masses ($m \neq m'$) of about $10^{16}$ GeV. If we assume that the relative mass splitting parameter $\delta m/m$ scales with temperature $T$ and it is of order unity initially at $T \sim 10^{16}$ GeV, then the temperature difference between the two sectors is reasonably understood and the tiny mass splitting parameter of $\delta m/m \sim 10^{-14}$ \cite{tan2019} for later quark condensation at $T\sim 10^{2}$ GeV is also explained.

The two scalar fields of $\phi$ and $\phi'$ then start to drive the inflation as two more space dimensions are extended. At the end of the space inflation, we have fully inflated the 4-d spacetime. Both $\phi$ and $\phi'$ will eventually decay into two sectors (i.e., ordinary and mirror sectors) of particles, respectively.

One important prediction from this inflation model is that the space dimensions undergo a double-inflation process and hence can present a signature of a dipole mode. This might explain the recent discovery of a large dipole component of cosmic acceleration in a reanalysis of type Ia supernova data by Colin \textit{et al.} \cite{colin2019}.

In 4-d spacetime, a gauge symmetry group of $U_f(6)\times SU_c(3)\times SU_w(2)\times U_Y(1)$ for ordinary particles and a similar mirror gauge group of $U_f(6)'\times SU_c(3)'\times SU_w(2)'\times U_Y(1)'$ for mirror particles are required by the supersymmetry principle. Here $U_f(6)$ stands for the gauged flavor symmetry (for six quark flavors) while the rest represent the symmetries of the well-studied Standard Model gauge interactions.

The new gauge symmetry provides bosons with 96 degrees of freedom in each sector and the corresponding gauge supersymmetry of SMM$^4$ asks for new Dirac fermion particles (massless quarks and leptons) with the same 96 degrees of freedom in each sector \cite{tan2019e,tan2020}. The space inflatons $\phi$ and $\phi'$ then decay into these new particles to reheat the universe and form the two ordinary and mirror sectors, respectively. The different energy scales of the two inflatons causes different temperatures ($T' < T$) during reheating in the two sectors that will satisfy the constraints of big bang nucleosynthesis \cite{kolb1985,hodges1993}. The $Z_2$ symmetry at this stage is the mirror symmetry with fully developed 4-d spacetime. Subsequently, the mirror symmetry and supersymmetry then undergo spontaneous symmetry breaking due to staged quark condensation at temperatures between $\sim 10^2$ and $\sim 0.1$ GeV eventually leading to the $N=4$ pseudo-SUSY theory SMM$^{4b}$ as shown in \hyperref[app:4b]{Appendix} \cite{tan2019e,tan2020}.
 
SMM$^{4b}$ (or a slight variant SMM$^{4\nu}$ due to possible neutrino condensation) is the model that governs the current Universe. Composite Higgs scalars due to quark condensation result in non-zero masses of all quarks and leptons that explicitly break supersymmetry (i.e., making it pseudo-SUSY). The hierarchy of their mass scales are determined by the energy scales emergent during the staged quark condensation process. In particular, neutrinos will acquire tiny masses commensurate with the mass differences between ordinary Higgs and mirror Higgs particles and hence exhibit oscillations between different neutrino flavors.

Even for this pseudo-SUSY theory, after considering the pseudo-Nambu-Goldstone bosons from the $U_f(6)$ symmetry breaking, the degrees of freedom between gauge bosons and matter fermions are balanced within each sector, i.e., $n_b=n_f=90$ as shown in Table \ref{tab:1} \cite{tan2019e,tan2020}. Note that the degrees of freedom are less than those in SMM$^{4}$ by six. It is because neutrino degeneracy (i.e., both sectors share the same set of neutrinos) reduces the degrees of freedom of fermions by six \cite{tan2019e}. Namely, left-handed neutrinos participate in gauge interactions of the ordinary sector only while right-handed neutrinos interact in the mirror sector only. Therefore, we obtain exactly the Standard Model's gauge group of $SU_c(3)\times SU_L(2)\times U_Y(1)$ (the $SU(2)$ weak interaction is left-handed only) for the ordinary sector and $SU_c(3)'\times SU_R(2)'\times U_Y(1)'$ for the right-handed mirror sector.

The above formalism eventually gives an immediate extension (SMM$^{4b}$) to the Standard Model that preserves all its goodness. In addition, it provides natural explanations for nature of neutrinos, generations of elementary particles, and choices of gauge interactions. Such a supersymmetric standard model with mirror matter extension can also solve many outstanding puzzles in fundamental physics. Most amazingly, it explains the observed dark energy scale of $10^{-3}$ eV assuming that gravitational vacuum energy is determined by a coherent sum of all scalar fields \cite{tan2019e}.

Along with predicted neutral hadron-mirror hadron oscillations, in particular, the two most important cases of $n-n'$ and $K^{0}-K^{0'}$ for neutrons and kaons, we can naturally and consistently understand many other enigmas such as neutron lifetime anomaly and dark matter \cite{tan2019}, baryon asymmetry of the universe \cite{tan2019c}, nucleosynthesis and evolution of stars \cite{tan2019a}, ultra-high energy cosmic rays \cite{tan2019b}, and unitarity of the CKM matrix \cite{tan2019d}. More importantly, new testable predictions and feasible experiments have been proposed and are relatively easy to carry out to directly verify or refute this new model SMM$^{4b}$ \cite{tan2019d}.

\section{Conclusions and Outlook}

A dynamic evolution of both the Universe and its underlying laws is presented. At different spacetime dimensions, the underlying theories are completely different with different interactions and different particles. There is no single unification theory for describing all energy scales at the same time. Intriguingly, all these different theories at different energy scales are connected with spontaneous symmetry breaking or phase transition processes.

This is actually in line with our experience in physics and other scientific research. New phenomena and new laws can emerge under phase transitions like superconductivity and other cases. The very nature of quantum indeterminism makes it necessary for pursuing studies of all sub-fields of physics. It may be also why we have so diversified scientific fields ranging from physics, chemistry, biology, to human brains. Even though they all seem to be based on the same physics, new laws in the new field could not be fully deduced from fundamental physics especially without knowing all the details of the possible ``phase transition''-like barriers in between the fields. Similar conclusions could be extended to economy, psychology, and other social science fields. In the end, our free will might be tied to indeterminism of quantum particles as indicated in the so-called free will theorem of Conway and Kochen \cite{conway2006,*conway2009}. Quantum indeterminism may be the fundamental reason that drives us to be open-minded, our research to follow open science practices, and our world to be open, collaborative, and prejudice-free to overcome the ``barriers'' between us.

\newpage

\bibliography{indeterm}

\newpage
\appendix*
\section{Supersymmetric mirror models\label{app}}
Here we list some technical details of supersymmetric mirror models (SMM) at various phases of the Universe and spacetime. See Refs. \cite{tan2019e,tan2020} for more details.

\vspace*{0.6\baselineskip}
\noindent\textbf{SMM$^{1}$:}\vspace*{0.4\baselineskip}

In 1-d time, SMM$^{1}$ provides the description of quantum fluctuations of massless true scalar field $\varphi$ with its Lagrangian and action as,
\begin{equation}\label{eq:l1dapp}
\mathcal{L} = \frac{1}{2}\dot{\varphi}^2, \qquad \mathcal{S} = \frac{1}{M^3_p}\int dt \mathcal{L}
\end{equation}
where the factor involving the Planck mass $M_p$ is obtained by a trivial integration over the three unextended or uninflated space dimensions.

\vspace*{0.6\baselineskip}\noindent\textbf{SMM$^{1b}$:}\vspace*{0.4\baselineskip}

To break this time reversal $Z_2$ symmetry, the spontaneous symmetry breaking mechanism is introduced by a potential term $V(\varphi) = -m^2\varphi^2/2+\varphi^4/8$ for the scalar $\varphi$ to acquire mass and then the SMM$^{1b}$ Lagrangian becomes
\begin{equation}\label{eq:l1dssb}
\mathcal{L} = \frac{1}{2}\dot{\varphi}^2 + \frac{1}{2}m^2\varphi^2 - \frac{1}{8}\varphi^4.
\end{equation}

\vspace*{0.6\baselineskip}\noindent\textbf{SMM$^{2}$:\label{app:2}}\vspace*{0.4\baselineskip}

In 2-d spacetime, the Lagrangian and the action are,
\begin{eqnarray}\label{eq:l2}
\mathcal{L} &=& -\frac{1}{4}F_{\mu\nu}F^{\mu\nu} + \frac{i}{2} \lambda^{\dagger} \bar{\sigma}^{\mu}\partial_{\mu} {\lambda} + h.c., \qquad
\mathcal{S} = \frac{1}{M^2_p}\int dtdx \mathcal{L}
\end{eqnarray}
where the $U(1)$ gauge field strength $F_{\mu\nu} = \partial_{\mu}A_{\nu} - \partial_{\nu}A_{\mu}$ and the massless Majorana fermion $\lambda$ has to be neutral and does not couple to the gauge field $A_{\mu}$. They form the simplest $N=1$ abelian gauge SUSY multiplet $(1,1/2)$ with the above on-shell Lagrangian. This could also be the model to describe the interior of a black hole studied under a separate work.

\vspace*{0.6\baselineskip}\noindent\textbf{SMM$^{2b}$:}\vspace*{0.4\baselineskip}

Spontaneous symmetry breaking will occur when Majorana fermions start to condense into two scalars of $\phi$ and $\phi'$. The $U(1)$ gauge symmetry is broken and the new Lagrangian at the new emerging energy scale of $10^{16}$ GeV is composed of two parts,
\begin{eqnarray}
\mathcal{L}_{(\frac{1}{2},0)} &=& \frac{1}{2}(\partial_{\mu}\phi\partial^{\mu}\phi + \partial_{\mu}\phi'\partial^{\mu}\phi') + \frac{i}{2} \lambda^{\dagger} \bar{\sigma}^{\mu}\partial_{\mu} {\lambda} + h.c. \\
\mathcal{L}_{\text{Higgs}} &=&  -(\lambda^{\dagger}_L\lambda_R(\phi+i\phi')+h.c.) + \frac{1}{2}m^2\phi^2 - \frac{1}{8}\phi^4 + \frac{1}{2}{m'}^{2}{\phi'}^{2} - \frac{1}{8}{\phi'}^{4}
\end{eqnarray}
where the first part shows the symmetry of $N=1$ pseudo-SUSY multiplet $(\frac{1}{2},0)$ and the second part presents a Higgs-like mechanism with masses ($m \neq m'$) of about $10^{16}$ GeV.

\vspace*{0.6\baselineskip}\noindent\textbf{SMM$^{4}$:}\vspace*{0.4\baselineskip}

In 4-d spacetime, gauge groups of $U_f(6)\times SU_c(3)\times SU_w(2)\times U_Y(1)$ for ordinary particles and $U_f(6)'\times SU_c(3)'\times SU_w(2)'\times U_Y(1)'$ for mirror particles are required by supersymmetry. The Lagrangian of the ordinary sector can then be written as,
\begin{eqnarray} \label{eq:LO}
\mathcal{L}_{\text{O}} &=& -\frac{1}{4}G^a_{\mu\nu}G^{a\mu\nu} + i\bar{\psi}_j \gamma^{\mu}D_{\mu} {\psi}_j 
\end{eqnarray}
where $G^a_{\mu\nu}$ ($a=1,2,...,48$) is the gauge field strength tensor and the gauge covariant derivative $D_{\mu} = \partial_{\mu} - i g T^a A^a_{\mu}$ depends on gauge symmetry generators $T^a$ and gauge bosons $A^a_{\mu}$. The massless Dirac fermion fields $\psi_j$ include three generations of all quarks and leptons. The degrees of freedom for gauge bosons and fermions are matched as $n_f = n_b = 96$ for the ordinary sector so that they can be in the same SUSY multiplet $(1,\frac{1}{2})$. Eq. (\ref{eq:LO}) as the on-shell Lagrangian therefore follows the unbroken $N=1$ supersymmetric Yang-Mills theory.

The mirror sector is essentially identical but completely decoupled from the ordinary sector regarding gauge interactions. Besides sharing the same gravity, i.e., the stage of spacetime, the two sectors are related under the mirror transformation $\mathcal{M}$ as follows,
\begin{equation}\label{eq_m1}
\mathcal{M}: \: \psi_L \rightarrow -\psi'_L, \: \psi_R \rightarrow \psi'_R, \: A_{\mu} \rightarrow A'_{\mu}
\end{equation}
which performs differently by a negative sign for left- and right-handed fermion fields \cite{tan2019e}.

\vspace*{0.6\baselineskip}\noindent\textbf{SMM$^{4b}$:\label{app:4b}}\vspace*{0.4\baselineskip}

Staged quark condensation occurs at energies between $10^2$ GeV - $10^2$ MeV along with breaking of the mirror symmetry and $U_f(6)$ as evidenced by the hierarchical quark and meson masses. The Lagrangian of the ordinary sector becomes
\begin{eqnarray} \label{eq:LOsm3}
\mathcal{L_{\text{O}}} &=& \mathcal{L}_{(1,\frac{1}{2})} + \mathcal{L}_{(\frac{1}{2},0)} + \mathcal{L}_{\text{Higgs}} \\
\mathcal{L}_{(1,\frac{1}{2})} &=& -\frac{1}{4}G^a_{\mu\nu}G^{a\mu\nu} + i\bar{\psi}^L_j \gamma^{\mu}D^L_{\mu} {\psi}^L_j + i\bar{\psi}^R_j \gamma^{\mu}D^R_{\mu} {\psi}^R_j \\
\mathcal{L}_{(\frac{1}{2},0)} &=& i\bar{\nu}^R_n \gamma^{\mu}\partial_{\mu} {\nu}^R_n + \frac{1}{2}(\partial_{\mu} \phi_f) (\partial^{\mu} \phi_f) \\
\mathcal{L}_{\text{Higgs}} &=& - \sum_j y_j(\bar{\psi}^L_j \psi^R_j\phi_f + h.c.) + \sum_f \frac{1}{2} m^2_f \phi_f^2 - \sum_f \frac{1}{8}(\phi_f)^4 \label{eq:LOhiggs}
\end{eqnarray}
where $\mathcal{L}_{(1,\frac{1}{2})}$ preserves an $N=1$ gauge pseudo-SUSY $(1,\frac{1}{2})$ multiplet, $\mathcal{L}_{(\frac{1}{2},0)}$ presents three copies of chiral SUSY $(\frac{1}{2},0)$ multiplets, and $\mathcal{L}_{\text{Higgs}}$ provides the mass and Higgs potential terms via the Higgs mechanism. Here the gauge group is reduced to $SU_c(3)\times SU_L(2)\times U_Y(1)$ with $SU_L(2)$ applying only to left-handed fermions. To keep both gauge bosons and fermions in the same $(1,\frac{1}{2})$ multiplet, the matching of DoFs is realized with $n_f = n_b = 90$ as pseudo-Nambu-Goldstone bosons from the $U_f(6)$ symmetry breaking provide 63 DoFs and neutrino degeneracy reduces DoFs of gauge-participating fermions by six \cite{tan2019e}. The three right-handed neutrinos ($\nu^R_n$, $n=1,2,3$) form the three chiral SUSY multiplets with the six real Higgs scalars ($\phi_f \sim \langle \bar{q}_f q_f\rangle$, $f=1,...,6$) from staged quark condensation \cite{tan2019e}. Therefore, overall the ordinary sector after the symmetry breaking shows a maximal $N=4$ pseudo-SUSY.

The mirror sector basically behaves the same obeying a similar $N=4$ pseudo-SUSY except that in this case the mirror gauge group is reduced to $SU_c(3)'\times SU_R(2)'\times U_Y(1)'$. The scalar fields are mirror-odd under the mirror symmetry transformation $\mathcal{M}:\: \phi \rightarrow -\phi'$.

Neutrino mass terms have to be constructed from both sectors as follows \cite{tan2019e},
\begin{equation}
-y(\bar{\nu}_L \nu_R \phi + \bar{\nu}'_L \nu'_R \phi' + h.c.) = -y(\bar{\nu}_L \nu_R (\phi-\phi') + h.c.)
\end{equation}
where masses of the neutrinos are determined by the ordinary-mirror mass splitting scale of $\langle \phi-\phi' \rangle \sim \delta v$ with fairly well constrained relative scale of $\delta v /v =10^{-15} \text{--} 10^{-14}$ \cite{tan2019,tan2019a,tan2019b,tan2019c,tan2019d,tan2019e}.

\end{document}